\documentclass{llncs}
\usepackage{cite}
\usepackage{amsmath}
\usepackage{mathtools}
\usepackage{amssymb}
\usepackage{amsfonts}
\let\proof\relax

\usepackage{amsthm}
\usepackage[linesnumbered,ruled,vlined]{algorithm2e}
\usepackage{graphicx}
\usepackage{textcomp}
\usepackage{enumitem}
\usepackage{url}
\usepackage{wrapfig}

\def\BibTeX{{\rm B\kern-.05em{\sc i\kern-.025em b}\kern-.08em
    T\kern-.1667em\lower.7ex\hbox{E}\kern-.125emX}}

\newtheorem{Definition}{Definition}
\newtheorem{exe}{Theorem}
\theoremstyle{remark}

\newcommand{\Comment}[1]{}
\newcommand{\sig}[1]{\textsf{{#1}\xspace}}
\newcommand{\keywords}[1]{\par\addvspace\baselineskip
\noindent\keywordname\enspace\ignorespaces#1}

\begin{document}

\title{ Verification of Binarized Neural Networks via Inter-Neuron Factoring 
}

\author{
Chih-Hong Cheng
\and
Georg N\"{u}hrenberg
\and
Chung-Hao Huang
\and
Harald Ruess}

\institute{
	fortiss - Landesforschungsinstitut des Freistaats Bayern\\
        Guerickestr. 25, 80805 Munich, Germany\\
        \vspace{1mm}
        \texttt{\{cheng,nuehrenberg,huang,ruess\}@fortiss.org}\\
}

\maketitle

\begin{abstract}

We study the problem of formal verification of Binarized Neural Networks (BNN), which have recently been proposed as a energy-efficient alternative to traditional learning networks. The verification of BNNs, using the reduction to hardware verification, can be even more scalable  by factoring computations among neurons within the same layer. By proving the NP-hardness of finding optimal factoring  as well as the hardness of PTAS approximability, we design polynomial-time search heuristics to generate factoring solutions. The overall framework allows applying verification techniques to moderately-sized BNNs for embedded devices with thousands of neurons and inputs. 

\end{abstract}

\section{Introduction}~\label{sec.introduction}

\vspace{-5mm}

Artificial neural networks have become essential building blocks in realizing many automated and even autonomous systems.
They have successfully been deployed, for example, for perception and scene understanding~\cite{krizhevsky2012imagenet,sermanet2013overfeat,long2015fully}, 
for control and decision making~\cite{chen2015deepdriving,huval2015empirical,sun2017fast,Lenz2017},  
and also for end-to-end solutions of autonomous driving scenarios~\cite{DBLP:journals/corr/BojarskiTDFFGJM16}\@.
Implementations of artificial neural networks, however, need to be made much more power-efficient in order to deploy them on typical embedded devices with  their
characteristically limited resources and power constraints.
Moreover, the use of neural networks in safety-critical systems poses severe verification
and certification challenges~\cite{bhattacharyya2015certification}\@.

Binarized Neural Networks (BNN) have recently been proposed~\cite{kim2016bitwise,courbariaux2016binarized} as a potentially much more power-efficient alternative to more traditional feed-forward artificial neural networks\@.
Their main characteristics are that trained weights, inputs, intermediate signals and outputs, and also activation constraints are binary-valued.
Consequently, forward propagation only relies on bit-level arithmetic\@. 
Since BNNs have also demonstrated good performance on standard datasets in image recognition such as MNIST, CIFAR-10 and SVHN~\cite{courbariaux2016binarized}, they are an attractive and potentially power-efficient alternative to current floating-point 
based implementations of neural networks for embedded applications.

In this paper we study the verification problem for BNNs.
Given a trained BNN and a specification of its intended input-output behavior, we develop 
verification procedures for establishing that the given BNN indeed meets its intended specification for all possible inputs.
Notice that naively solving verification problems for BNNs with, say, $1000$ inputs requires investigation of all $2^{1000}$ different input configurations.

For solving the verification problem of BNNs we build on well-known methods and tools from the hardware verification domain. 
We first transform the BNN and its specification into a {\em combinational miter}~\cite{brayton2010abc}, which  is then transformed into a corresponding propositional satisfiability (SAT) problem. 
In this process we rely heavily on logic synthesis tools such as~\texttt{ABC}~\cite{brayton2010abc} from the hardware verification domain\@. 
Using such a direct neuron-to-circuit encoding, however, we were not able to verify BNNs with thousands of inputs and hidden nodes, as encountered in some of our embedded systems case studies.
The main challenge therefore is to make the basic verification procedure scale to BNNs as used on current embedded devices. 
 
It turns out that one critical ingredient for efficient BNN
verification is to factor computations among neurons in the same layer, which
is possible due to weights being binary. Such a technique is not applicable
within recent works in verification of floating point neural
networks~\cite{pulina2010abstraction,DBLP:conf/cav/KatzBDJK17,cheng2017maximum,ehlers2017formal,lomuscio2017approach}.
The key theorem regarding the hardness of finding optimal factoring as well as
the hardness of inapproximability 
leads to the design of polynomial time search heuristics for generating
factorings. These factorings substantially increase the scalability of formal
verification via SAT solving.

The paper is structured as follows.
Section~\ref{sec.bnn}  defines basic notions and concepts underlying BNNs.
Section~\ref{sec.formal.verification} presents our verification workflow including the factoring of counting
units (Section~\ref{sub.sec.counting})\@. 
We summarize experimental results with our verification procedure in Section~\ref{sec.evaluation}, 
compare our results with related work from the literature in Section~\ref{sec.related}, 
and we close with some final remarks and an outlook in Section~\ref{sec.conclusion}\@. Proofs of theorems are listed in the appendix.

\begin{table}[t]
	
	\centering
	\setlength\doublerulesep{0.2cm} 
	\begin{small}
		\begin{tabular}{|l|c|c|c|c|c|}
			\hline
			index j   & 0 (bias node) & 1 & 2 & 3 & 4    \\
			\hline 
			$x^{(l-1)}_j$   & +1 (constant) & +1 & -1 & +1 & +1       \\
			\hline 
			$w^{(l)}_{ji}$   & -1  (bias) & +1 & -1 & -1 & +1       \\
			\hline 
			$x^{(l-1)}_j w^{(l)}_{ji}$   & -1 & +1 & +1 & -1 & +1       \\
			\hline 
			$\textsf{im}^{(l)}_i$ & \multicolumn{5}{c|}{{\scriptsize$(-1) + ( +1 ) + (+1) + (-1) + (+1) = 1$}} \\ 
			\hline 
			$\textsf{x}^{(l)}_i$ & \multicolumn{5}{c|}{+1, as $\textsf{im}^{(l)}_i > 0$ } \\ 
			\hline 
			\hline 
			index j   & 0 (bias node) & 1 & 2 & 3 & 4    \\
			\hline 
			$x^{(l-1)}_j$   & \sig{1} & \sig{1} & \sig{0} &  \sig{1} & \sig{1}       \\
			\hline 
			$w^{(l)}_{ji}$   & \sig{0}  (bias) &  \sig{1} & \sig{0} & \sig{0} & \sig{1}       \\
			\hline 
			$x^{(l-1)}_j\, \overline{\oplus}\, w^{(l)}_{ji}$   &  \sig{0} &  \sig{1} &  \sig{1} & \sig{0} &  \sig{1}       \\
			\hline 
			\# of  \sig{1}'s in $x^{(l-1)}_j\, \overline{\oplus}\, w^{(l)}_{ji}$ & \multicolumn{5}{c|}{3} \\ 
			\hline 
			$\textsf{x}^{(l)}_i$ & \multicolumn{5}{c|}{$\sig{1}$, as $(3\geq \lceil\frac{5}{2}\rceil)$} \\ 
			\hline 
		\end{tabular}			
	\end{small}
	\vspace{1mm}
	\caption{An example of computing the output of a BNN neuron, using bipolar domain (up) and using \sig{0}/\sig{1} boolean variables (down).}
	\label{table.xnor.computation.example}
\end{table}

\begin{figure}[t]
	\centering
	\includegraphics[width=0.6\columnwidth]{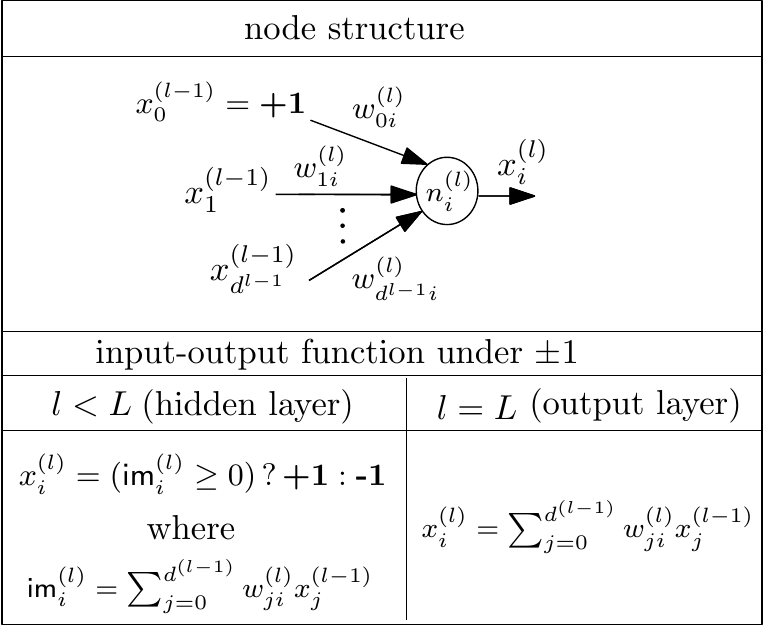}	
	\caption{Computation inside a neuron of a BNN, under bipolar domain $\pm 1$.}
	\label{fig:neuralnetwork.example}
\end{figure}

\section{Preliminaries}~\label{sec.bnn}
\vspace{-5mm}

Let $\mathbb{B}$ be the set of \emph{bipolar binaries} $\pm 1$,  where +1 is interpreted as “\textsf{true}” and $-1$ as “\textsf{false}.”
A \emph{Binarized  Neural Network} (BNN)~\cite{kim2016bitwise,courbariaux2016binarized} consists of a sequence of layers labeled from $l=0,1, \ldots, L$, where~$0$ is the index of the \emph{input layer},~$L$ is the \emph{output layer}, and all other layers are so-called \emph{hidden layers}\@.
Superscripts~$^{(l)}$ are used to index layer~$l$-specific variables. 
Elements of both inputs and outputs vectors of a BNN are of bipolar domain $\mathbb{B}$\@.

Layers~$l$ are comprised of \emph{nodes} $n^{(l)}_i$ (so-called neurons), for
$i=0,1, \ldots, d^{(l)}$, where~$d^{(l)}$ is the {\em dimension} of the layer~$l$. 
By convention, $n_{0}^{(l)}$ is a \emph{bias node} and has constant bipolar output~$+1$.
Nodes~$n^{(l-1)}_j$ of layer~$l-1$ can be connected with nodes~$n^{(l)}_i$ in layer~$l$ by 
a directed edge of \emph{weight} $w^{(l)}_{ji}\in \mathbb{B}$. 
A layer is fully connected if every node (apart from the bias node) in the layer is connected to all neurons in the previous layer. 
Let
$\vec{w}^{(l)}_{i}$ denote the array of all
weights associated with neuron $n^{(l)}_i$. Notice that we consider all
weights in a network to have fixed bipolar values.

Given an input to the network, computations are applied successively from
neurons in layer $1$ to $L$ for generating outputs.
Fig.~\ref{fig:neuralnetwork.example} illustrates the computations of a neuron
in bipolar domain. Overall, the activation
function is applied to the intermediately computed weighted sum. It  outputs~$+1$ if the
weighted sum is greater or equal to~$0$; otherwise, output~$-1$. For the output layer, the activation function is omitted. 
For~$l = 1,\dots, L$ let $x^{(l)}_i$ denote the output value of node
$n^{(l)}_i$ and $\vec{x}^{(l)} \in \mathbb{B}^{|d^{(l)}|+1}$ denotes the array
of all outputs from layer~$l$, including the constant bias node;
$\vec{x}^{(0)}$ refers to the input layer.

For a given BNN and a relation $\phi_{risk}$ specifying the undesired property between the bipolar input and output domains of the given BNN, the {\em BNN safety verification problem} asks if there exists an input $\vec{a}$ to the BNN such that the risk property $\phi_{risk}(\vec{a},\vec{b})$ holds, where $\vec{b}$ is the output of the BNN for input~$\vec{a}$\@.

It turns out that safety verification of BNN is no simpler than safety verification of floating point neural networks with ReLU activation function~\cite{DBLP:conf/cav/KatzBDJK17}. Nevertheless, compared to floating point neural networks, the simplicity of binarized weights allows an efficient translation into SAT problems, as can be seen in later sections.

\begin{exe}
	The problem of BNN safety verification is NP-complete. 
\end{exe}

\section{Verification of BNNs via Hardware Verification}~\label{sec.formal.verification}
\vspace{-2mm}

The BNN verification problem  is encoded by means of a \emph{combinational miter}~\cite{brayton2010abc}, which is a hardware circuit with only one Boolean output and the output should always be \sig{0}\@.
The main step of this encoding is to replace the  bipolar domain operation in
the definition of BNNs with corresponding operations in the \sig{0/1} Boolean domain.

We recall the encoding of the update function of an individual neuron of a BNN
in bipolar domain (Eq.\@~\ref{eq.xnor.counting.encoding}) by means of operations 
in the \sig{0/1} Boolean domain~\cite{kim2016bitwise,courbariaux2016binarized}:
(1) perform a bitwise XNOR ($\overline{\oplus}$) operation,  (2) count the number of
\sig{1}s, and (3) check if the sum is greater than or equal to the half
of the number of inputs being connected.
Table~\ref{table.xnor.computation.example} illustrates the concept by providing
the detailed computation for a neuron connected to five predecessor nodes. 
Therefore, the update function of a BNN neuron (in the fully connected layer) in the Boolean domain is as follows. 

\begin{equation}~\label{eq.xnor.counting.encoding}
x^{(l)}_i = \textsf{geq}_{\big\lceil\frac{|d^{(l-1)}|+1}{2}\big\rceil }(\textsf{count1}(\vec{w}^{(l)}_{i} \, \overline{\oplus}\, \vec{x}^{(l-1)}))\mbox{~,} 
\end{equation}

where $\textsf{count1}$ simply counts the number of \sig{1}s in an array of
Boolean variables, and
$\textsf{geq}_{\big\lceil\frac{|d^{(l-1)}|+1}{2}\big\rceil }(x)$ is $\sig{1}$
if $x \geq \big\lceil\frac{|d^{(l-1)}|+1}{2}\big\rceil$, and $\sig{0}$ otherwise. 
Notice that the value $\big\lceil\frac{|d^{(l-1)}|+1}{2}\big\rceil$ is constant for a given BNN. 

Specifications in the bipolar domain can also be easily re-encoded in the Boolean domain.
	Let $(x_{i}^{(L)})_{_{\pm1}}$ be the valuation in the bipolar domain and $(x_{i}^{(L)})_{_{\textsf{0/1}}}$ be the output valuation in the Boolean domain; then the transformation from bipolar to Boolean domain is as follows.

	\begin{equation}
	(x_{i}^{(L)})_{_{\pm1}} = 2\cdot(x_{i}^{(L)})_{_{\textsf{0/1}}} - d^{(L-1)}
	\label{eq.xnor.counting.id}
	\end{equation}
	An illustrative example is provided in Table~\ref{table.xnor.computation.example}, where $\sig{im}^{(l)}_{i}=1 = 2\cdot 3 - 5$\@.
	In the remaining of this paper we assume that properties are always provided 
	in the Boolean domain.

\subsection{From BNN to hardware verification}~\label{sub.sec.miter.encoding}
\vspace{-2mm}

We are now ready for stating the basic decision procedure for solving BNN
verification problems. This procedure first constructs a combinational miter
for a BNN verification problem, followed by an encoding of the combinational
miter into a corresponding propositional SAT problem. Here we rely on standard
transformation techniques as implemented in logic synthesis tools such as
\texttt{ABC}~\cite{brayton2010abc} or \texttt{Yosys}~\cite{wolf2013yosys} for
constructing SAT problems from miters.
The decision procedure takes as input a BNN network description, an
input-output specification $\phi_{risk}$ and can be summarized by the following workflow:

\begin{enumerate}
	\item Transform all neurons of the given BNN into
	neuron-modules. All neuron-modules have identical structure,
	but only differ based on the associated weights and biases of the
	corresponding neurons.
	\item Create a BNN-module by wiring the neuron-modules realizing the
	topological structure of the given BNN.
	\item Create a property-module for the property $\phi_{risk}$. Connect
	the inputs of this module with all the inputs and all the
	outputs of the BNN-module. The output of this module is \sig{true} if
	the property is satisfied and \sig{false} otherwise.
	\item The combination of the BNN-module and the property-module is the \sig{miter}.
	\item Transform the \sig{miter} into a propositional SAT formula.
	\item Solve the SAT formula. If it is unsatisfiable then the BNN is safe w.r.t.\
	$\phi_{risk}$; if it is satisfiable then the BNN exhibits the risky behavior
	being specified in $\phi_{risk}$.

\end{enumerate}

\subsection{Counting optimization}~\label{sub.sec.counting}
\vspace{-2mm}

\begin{figure}[t]
	\centering
	\includegraphics[width=0.35\columnwidth]{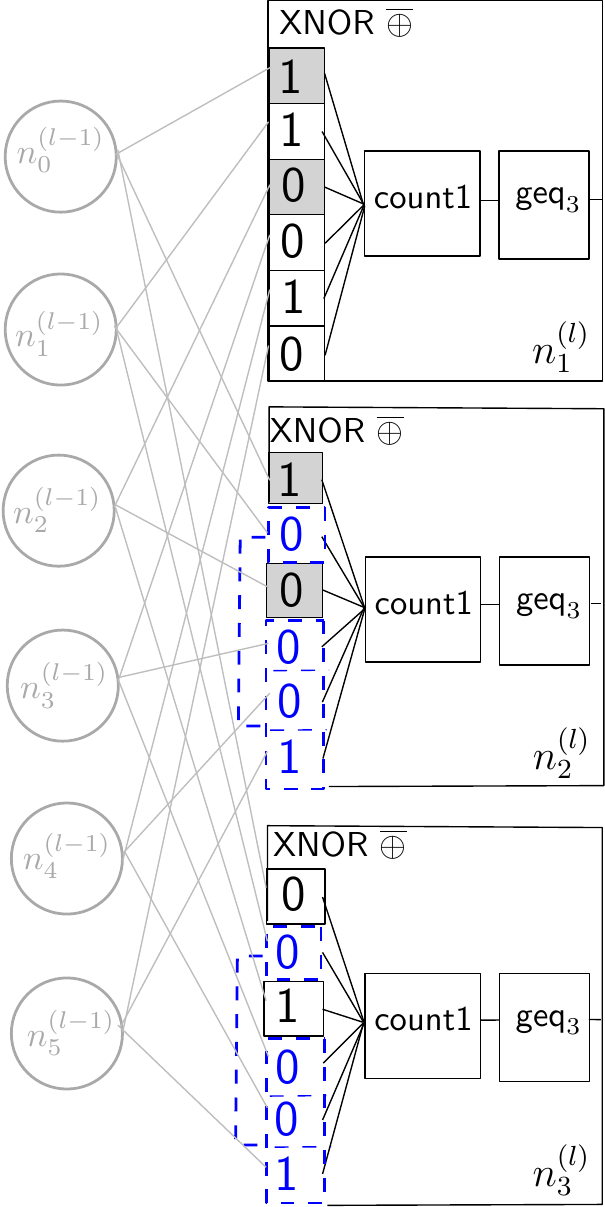}	
	\caption{One possible factoring to avoid redundant counting.}
	\label{fig:count.factoring}
\end{figure}

The goal of the counting optimization is to speed up SAT-solving times by
reusing redundant counting units in the circuit and, thus, reducing redundancies
in the SAT formula. This method involves the identification and factoring of
redundant counting units, illustrated in Figure~\ref{fig:count.factoring},
which highlights one possible factoring.
The main idea is to exploit similarities among the weight vectors of neurons in
the same layer, because the counting over a portion of the weight vector has
the same result for all neurons that share it. The circuit size is reduced by
using the factored counting unit in multiple neuron-modules. We define a
factoring as follows:

\Comment{
	\begin{algorithm}[t]
		\label{algorithm.bnn.verification}
		\KwData{BNN network description (see Section~\ref{sec.bnn}) and an input-output spec $\phi_{risk}$}
		\KwResult{Whether there exists an input $\vec{a}$ such that $\phi_{risk}(\vec{x}^{(0)}(\vec{a}), \vec{x}^{(L)}(\vec{a}))$ holds}
		
		$\sig{miter}.\sig{inputSize} = d^{(0)}$, $\sig{miter}.\sig{outputSize} = 1$\;	
		\lForEach{neuron $n^{(l)}_i$, $l,i>0$}{
			$\sig{module}[n^{(l)}_i]  := \sig{build\_neuron\_module}(\vec{w}^{(l-1)}_i)$\;
			$\sig{miter}.\sig{modules} = \sig{miter}.\sig{modules} \cup \{\sig{module}[n^{(l)}_i]\}$
		}
		\lForEach{neuron $n^{(l)}_i$, $n^{(l-1)}_j$, $l>1,i\geq 0$}{
			$\sig{miter}.\sig{add\_wire}(\sig{module}[n^{(l-1)}_j].\sig{out}, \sig{module}[n^{(l)}_i].\sig{in}_j)$
		}
		\lForEach{neuron $n^{(l)}_i$, $l=1,i>0$}{
			$\sig{miter}.\sig{add\_wire}(\sig{miter}.\sig{in}_j, \sig{module}[n^{(l)}_i].\sig{in}_j)$
		}
		\lForEach{neuron $n^{(l)}_i$, $l\geq 1$}{
			$\sig{miter}.\sig{add\_wire}($\sig{constant(1)}$, \sig{module}[n^{(l)}_0].\sig{in}_j)$
		}
		$\sig{miter}.\sig{out} := \phi_{risk}(  \sig{miter}.\sig{in} \mapsto \vec{x}^{(0)},   (\sig{module}[n^{(L)}_0].\sig{out}, \ldots, \sig{module}[n^{(L)}_{d^{(L)}}].\sig{out}) \mapsto \vec{x}^{(L)})$\;
		
		$\Sigma_{BNN} := \sig{generate\_CNF\_constraint}(\sig{miter})$\;
		
		\leIf{$\Sigma_{BNN}$ is SAT}{\Return \sig{true}}{\Return \sig{false}}
		
		\caption{Decision procedure for BNN verification problems.}
		
	\end{algorithm}
}

\begin{Definition}[factoring and saving] Consider the $l$-th layer of a BNN where  $l>0$.
	A \emph{factoring} $f = (I, J)$ is a pair of two sets, where $I\subseteq
	\{1,\dots,d^{(l)}\}$, $J\subseteq \{1,\dots,d^{(l-1)}\}$, such that $|I|>1$, and for all  $i_1, i_2\in I$, for all $j \in J$, we have $w^{(l)}_{j i_1} = w^{(l)}_{ j i_2}$. Given a factoring $f = (I, J)$, define its \emph{saving} $\emph{\sig{sav}}(f)$ be $(|I|-1)\cdot|J|$. 
	
\end{Definition}

\begin{Definition}[non-overlapping factorings]
	Two factorings $f_1= (I_1, J_1)$ and $f_2 = (I_2, J_2)$ are
	non-overlapping when the following condition folds: if $(i_1, j_1) \in f_1$ and $(i_2, j_2) \in f_2$, then either $i_1 \neq i_2$ or $j_1 \neq j_2$. In other words, weights associated with $f_1$ and $f_2$ do not overlap.
\end{Definition}

\begin{Definition}[$k$-factoring optimization problem]
	The $k$-factoring optimization problem searches for a set $F$ of size~$k$
	factorings $\{f_1, \ldots, f_k\}$, such that any two factorings are
	non-overlapping, and the total saving $\emph{\sig{sav}}(f_1) + \dots +
	\emph{\sig{sav}}(f_k)$  is maximum. 
	
\end{Definition}

\noindent For the example in Fig.~\ref{fig:count.factoring}, there are two non-overlapping factorings $f_1 = (\{1,2\}, \{0,2\})$ and 
$f_2 = (\{2,3\}, \{1,3,4,5\})$.  $\{f_1, f_2\}$ is also an optimal solution for the $2$-factoring optimization problem, with the total saving being~$(2-1)\cdot 2+ (2-1)\cdot4 = 6$. Even finding one factoring $f_1$ which has the overall maximum saving $\sig{sav}(f_1)$, is computationally hard. This NP-hardness result is established by a reduction from the NP-complete problem of finding maximum edge biclique in bipartite graphs~\cite{peeters2003maximum}.

\begin{exe}[Hardness of factoring optimization]
	The $k$-factoring optimization problem, even when $k=1$,  is NP-hard.
\end{exe}

\noindent Furthermore, even having an approximation
algorithm for the $k$-factoring optimization problem is hard - there is no polynomial time approximation scheme (PTAS), unless NP-complete problems can be solved in randomized subexponential time. The proof follows an intuition that building a PTAS for $1$-factoring can be used to build a PTAS for finding maximum complete bipartite subgraph which also has known inapproximability results~\cite{ambuhl2011inapproximability}.

\begin{exe}
	Let $\epsilon > 0$ be an arbitrarily small constant. If there is a PTAS
	for the $k$-factoring optimization problem, even when $k=1$, then there
	is a (probabilistic) algorithm that decides whether a given SAT
	instance of size $n$ is satisfiable in time~$2^{n^{\epsilon}}$.	
\end{exe}

\SetArgSty{textrm}
\begin{algorithm}[t]
	\SetInd{0.5em}{0.4em}
	\begin{footnotesize}
		\SetKwProg{Fn}{Function}{ is}{end}
		\label{algorithm.counting.heuristic}
		\KwData{BNN network description (cf Sec.~\ref{sec.bnn}) }
		\KwResult{Set $F$ of factorings, where any two factorings of $F$ are non-overlapping. } 	
		\SetKwFunction{FMain}{main}
		\SetKwFunction{FGetMaxSaving}{getFactoring}
		\SetKwProg{Fn}{function}{:}{}
		\Fn{\FMain{}}{
			\textbf{let} $\sig{used} := \emptyset$ and $F:= \emptyset$\; 
			
			\ForEach{neuron $n^{(l)}_i$}{
				\textbf{let} $\sig{f}^{opt}_i :=$ empty factoring\; 
				\ForEach{weight $w^{(l)}_{ji}$ where $(i,j) \not\in \sig{used}$}{
					$f_{ij} = \texttt{getFactoring}(i, j,
					\sig{used})$\;
					\lIf{$\sig{sav}(f_{ij}) > \sig{sav}(f^{opt}_i)$}{$f^{opt}_i := f_{ij}$}
				}
				$\sig{used} := \sig{used} \cup \{ (i,j)\; |\; (i, j)\in f^{opt}_i\}$; 
				$F := F \cup \{f^{opt}_i\}$\;
			}
			\KwRet $F$\;
		}
		
		\SetKwProg{Pn}{function}{:}{\KwRet}
		\Pn{\FGetMaxSaving{$i$, $j$, \sig{used}}}{
			\textbf{build}  $\mathbb{I}$ := $\{I_0,...,I_{d^{(l-1)}}\}$ where $I_{j'}$ := $\{i'\in\{0,...,d^{(l)}\}\ \big{|}\ w^{(l)}_{j'i'}=w^{(l)}_{j'i}\wedge(i',j')\not\in\sig{used} \}$\;
			\lForEach {$I_m\in\mathbb{I}$}{ $I_m$ := $I_m\bigcap I_j$}	
			\textbf{build}  $\mathbb{J}$ := $\{J_0,\ldots, J_{j'},\ldots,J_{d^{(l-1)}}\}$ where $J_{j'}$ := $\{j''\in\{0,...,d^{(l-1)}\}\ \big{|}\ I_{j'}\subseteq I_{j''}\}$\;	
			\KwRet $(I, J) := (I_{j^{*}},J_{j^{*}})$ where $I_{j^{*}}\in \mathbb{I},J_{j^{*}} \in \mathbb{J}$, and $(|I_{j^{*}}| - 1)\cdot|J_{j^{*}}| = 
			\sig{max}_{j'\in\{0,...,d^{(l-1)}\}}\ (|I_j'| - 1)\cdot|J_j'|\ $\;
		}	
	\end{footnotesize}

	
	\caption{Finding factoring possibilities for BNN.}
\end{algorithm}

\noindent As finding an optimal factoring is computationally hard, we  present a
\emph{polynomial time heuristic
	algorithm} (Algorithm~\ref{algorithm.counting.heuristic}) that finds factoring
possibilities among neurons in layer~$l$. The \texttt{main} function searches
for an unused pair of neuron $i$ and input $j$~(line~3~and~5), considers a
certain set of factorings determined by the subroutine
\texttt{getFactoring}~(line~6) where weight $w_{ji}^{(l)}$ is guaranteed to be used (as input parameter $i$, $j$), picks the factoring with greatest
$\sig{sav}()$~(line~7) and then adds the factoring greedily and updates the set
$\sig{used}$~(line~8).

\begin{figure}[t]

	\centering
	\includegraphics[width=0.8\columnwidth]{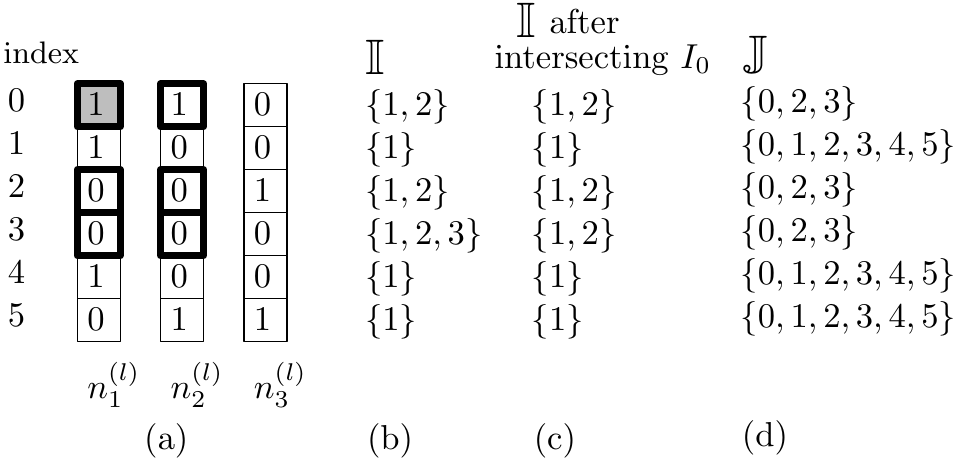}	

	\caption{Executing $\texttt{getFactoring}(1, 0, \emptyset)$, meaning that we consider a factoring which includes the top-left corner of (a). The returned factoring is highlighted in thick lines. }	
	\label{fig:heuristic.factoring}
\end{figure}

The subroutine $\texttt{getFactoring}()$ (lines 10--14) computes a factoring $(I,J)$ guaranteeing that weight $w_{ji}^{(l)}$ is used. It starts by creating a set~$\mathbb{I}$, where each element $I_{j'} \in \mathbb{I}$
is a set containing the indices of neurons whose $j'$-th weight matches the $j'$-th weight in neuron~$i$ (the condition $(w^{(l)}_{j'i'} = w_{j'i}^{(l)})$ in line 11). In the example in Fig.~\ref{fig:heuristic.factoring}a, the computation generates Fig.~\ref{fig:heuristic.factoring}b where $I_{3} = \{1,2,3\}$ as $w_{31}^{(l)} = w_{32}^{(l)} = w^{(l)}_{33} = 0$. The intersection performed on line~12 
guarantees that the set $I_{j'}$ is always a subset of $I_j$ -- as weight $w_{ji}$ should be included,  $I_j$ already defines the maximum set of neurons where factoring can happen. 
E.g., $I_{3}$ changes from $\{1,2,3\}$ to $\{1,2\}$ in Fig.~\ref{fig:heuristic.factoring}c. 

The algorithm then
builds a set $\mathbb{J}$ of all the candidates for $J$. Each element $J_{j'}$
contains all the inputs $j''$ that would benefit from $I_{j'}$ being 
the final result $I$. Based on the observation mentioned above, $J_{j'}$ can be
built through superset computation between elements of $\mathbb{I}$ (line~13,
Fig.~\ref{fig:heuristic.factoring}d). After we build $\mathbb{I}$ and~$\mathbb{J}$, finally line~14 finds a pair of $(I_{j^{*}},J_{j^{*}})$ where
$I_{j^{*}}\in \mathbb{I}$, $J_{j^{*}}\in \mathbb{J}$ with the maximum saving
$(|I_j^{*}| - 1)\cdot|J_j^{*}|$. The maximum saving as produced in
Fig.~\ref{fig:heuristic.factoring} equals $(|\{1,2\}|-1)\cdot |\{0,2,3\}| = 3$.

There are only polynomial operations in this algorithm such as nested for loops, superset checking and intersection  which makes the heuristic algorithm polynomial. When one encounters a huge number of neurons and long weight vectors, we further partition neurons and weights into smaller regions as input to  Algorithm~\ref{algorithm.counting.heuristic}. By doing so, we find factoring possibilities for each weight segment of a neuron and the algorithm can be executed in parallel.

\vspace{-3mm}
\section{Implementation and Evaluation}~\label{sec.evaluation}

\vspace{-5mm}

We have created a verification tool, which first reads a BNN description based on
the Intel Nervana Neon framework\footnote{
	\url{https://github.com/NervanaSystems/neon/tree/master/examples/binary}
}, generates a combinational miter in Verilog and calls
\texttt{Yosys}~\cite{wolf2013yosys} and \texttt{ABC}~\cite{brayton2010abc} for
generating a CNF formula. 
No further optimization commands (e.g., \texttt{refactor}) are executed inside
\texttt{ABC} to create smaller CNFs\@. 
Finally, \texttt{Cryptominisat5}~\cite{soos2016cryptominisat} is used for solving SAT queries. The experiments are conducted in a Ubuntu 16.04 Google Cloud VM equipped with 18 cores and 250 GB RAM, with \texttt{Cryptominisat5} running with 16 threads. We use two different datasets, namely the MNIST dataset for digit recognition~\cite{lecun1998mnist
} and the German traffic sign dataset~\cite{Stallkamp-IJCNN-2011}. We binarize the gray scale data to $\pm 1$ before actual training. 
For the traffic sign dataset, every pixel is quantized to~$3$ Boolean variables.

\begin{table}[t]
	\centering
	\begin{scriptsize}
		\begin{tabular}{|c|p{0.8cm}|p{1.5cm}|p{4cm}|p{1.2cm}|p{1.9cm}|p{1.9cm}| }
			\hline
			ID	&  \# inputs &  \#  neurons hidden layer  & Properties being investigated & SAT/ \newline UNSAT & SAT solving  time (normal) & SAT solving  time (factored)   \\ \hline
			MNIST 1	& 784 & 3x100 & $\textsf{out}_1 \geq 18 \wedge  \textsf{out}_2 \geq 18$   ($\geq18\%$)& SAT &  2m16.336s & 0m53.545s \\ \hline
			MNIST 1	& 784 & 3x100 & $\textsf{out}_1 \geq 30 \wedge  \textsf{out}_2 \geq 30$  ($\geq30\%$) & SAT & 2m20.318s & 0m56.538s \\ \hline
			MNIST 1	& 784 & 3x100 & $\textsf{out}_1 \geq 60 \wedge  \textsf{out}_2 \geq 60$  ($\geq60\%$) & SAT & timeout & 10m50.157s \\ \hline
			MNIST 1	& 784 & 3x100 & $\textsf{out}_1 \geq 90 \wedge  \textsf{out}_2 \geq 90$  ($\geq90\%$) & UNSAT & 2m4.746s  & 1m0.419s \\ \hline\hline
			Traffic 2	& 2352 & 3x500 & $\textsf{out}_1 \geq 90 \wedge  \textsf{out}_2 \geq 90$   ($\geq18\%$)& SAT & 10m27.960s & 4m9.363s \\ \hline
			Traffic 2	& 2352 & 3x500 & $\textsf{out}_1 \geq 150 \wedge  \textsf{out}_2 \geq 150$   ($\geq30\%$)& SAT & 10m46.648s & 4m51.507s \\ \hline
			Traffic 2	& 2352 & 3x500 & $\textsf{out}_1 \geq 200 \wedge  \textsf{out}_2 \geq 200$  ($\geq40\%$) & SAT & 10m48.422s & 4m19.296s \\ \hline
			Traffic 2	& 2352 & 3x500 & $\textsf{out}_1 \geq 300 \wedge  \textsf{out}_2 \geq 300$  ($\geq60\%$) & unknown & timeout & timeout  \\ \hline		
			Traffic 2	& 2352 & 3x500 & $\textsf{out}_1 \geq 475 \wedge  \textsf{out}_2 \geq 475$  ($\geq95\%$) & UNSAT & 31m24.842s & 41m9.407s \\ \hline	\hline				
			Traffic 3	& 2352 & 3x1000 & $\textsf{out}_1 \geq 120 \wedge  \textsf{out}_2 \geq 120$  ($\geq12\%$) & SAT & out-of-memory & 9m40.77s \\ \hline
			Traffic 3	& 2352 & 3x1000 & $\textsf{out}_1 \geq 180 \wedge  \textsf{out}_2 \geq 180$   ($\geq18\%$)& SAT & out-of-memory & 9m43.70s  \\ \hline
			Traffic 3	& 2352 & 3x1000 & $\textsf{out}_1 \geq 300 \wedge  \textsf{out}_2 \geq 300$  ($\geq30\%$) & SAT & out-of-memory & 9m28.40s \\ \hline
			Traffic 3	& 2352 & 3x1000 & $\textsf{out}_1 \geq 400 \wedge  \textsf{out}_2 \geq 400$  ($\geq40\%$) & SAT & out-of-memory & 9m34.95s \\ \hline
			
		\end{tabular}
	\end{scriptsize}

	\caption{Verification results for each instance and comparing the
		execution times of the plain hardware verification approach and the
		optimized version using counting optimizations.} 	
	\label{tab:monolitic}	
	
\end{table}

Table~\ref{tab:monolitic} summarizes the result of verification in terms of SAT
solving time, with a timeout set to~$90$ minutes\@.
The properties that we use here are characteristics of a BNN given by numerical
constraints over outputs, such as ``simultaneously classify an image as a
priority road sign and as a stop sign with high confidence'' (which clearly demonstrates a risk behavior).  
It turns out that factoring techniques are essential to enable better
scalability, as it halves the verification times in most cases and enables us
to solve some instances where the plain approach ran out of memory or timed
out. 
However, we also observe that solvers like \texttt{Cryptominisat5} might get
trapped in some very hard-to-prove properties.
Regarding the instance in Table~\ref{tab:monolitic} where the result is
unknown, we suspect that the simultaneous confidence value of $60\%$ for the
two classes $\textsf{out}_1$ and $\textsf{out}_{2}$, is close to the value
where the property flips from satisfiable to unsatisfiable. This makes SAT
solving on such cases extremely difficult for solvers as the instances are
close to the ``border'' between SAT and UNSAT instances.

Here we omit technical details, but the counting approach can also be replaced by techniques such as sorting networks\footnote{Intuitively, the counting + activation function can be replaced by implementing a sorting network and check if for the $m$ sorted result, the $\frac{m}{2}$-th element is~\sig{true}.}~\cite{sortingnetwork} where the technique of factoring can still be integrated\footnote{Sorting network in~\cite{sortingnetwork} implements merge-sort in hardware, where the algorithm tries to build a sorted string via merging multiple sorted substrings. Under the context of BNN verification, the factored result can be first sorted, then these sorted results can then be integrated as an input to the merger.}. However, our initial evaluation demonstrated that using sorting network does not bring any computational benefit.

\vspace{-2mm}
\section{Related Work}~\label{sec.related}
\vspace{-4mm}

There has been a flurry of recent results on formal verification of neural
networks  (e.g.~\cite{pulina2010abstraction,DBLP:conf/cav/KatzBDJK17,cheng2017maximum,ehlers2017formal,lomuscio2017approach})\@. 
These approaches usually  target the formal verification of floating-point arithmetic neural networks (FPA-NNs)\@. 
Huang et al.\ propose an (incomplete) search-based technique based on {\em satisfiability modulo theories} (SMT) solvers~\cite{DBLP:conf/cav/HuangKWW17}\@. 
For FPA-NNs with ReLU activation functions, Katz et al.\ propose a modification
of the Simplex algorithm which prefers fixing of binary
variables~\cite{DBLP:conf/cav/KatzBDJK17}\@. This verification approach has been
demonstrated on the verification of a collision avoidance system for UAVs.
In our own previous work on neural network verification we establish maximum resilience bounds for FPA-NNs based on reductions to {\em mixed-integer linear programming} (MILP) problems~\cite{cheng2017maximum}\@.  The feasibility of this approach has work has demonstrated, for example, by verifying a motion predictor in a highway overtaking scenario. 
The work of Ehlers~\cite{ehlers2017formal} is based on sound abstractions, and
approximates non-linear behavior in the activation functions. Scalability is
the overarching challenge for these formal approaches to the verification of
FPA-NNs. Case studies and experiments reported in the literature are usually
restricted to the verification of FPA-NNs with a couple of hundred neurons.

Around the time (Oct 9th, 2017) we first release of our work regarding formal verification of BNNs, Narodytska et al have also worked on the same problem~\cite{nina2017bnn}. Their work focuses on efficient encoding within a single neuron, while we focus on computational savings among neurons within the same layer. One can view our result and their result complementary.

Researchers from the machine learning domain (e.g. ~\cite{goodfellow2014generative,goodfellow2014explaining,moosavi2016universal})  target the generation of adversarial examples for debugging and retraining purposes.
Adverserial examples are slightly perturbed inputs (such as images) which may fool a neural network into generating undesirable results (such as "wrong" classifications)\@. 
Using satisfiability assignments from the SAT solving stage in our verification procedure, we are also able to generate {\em counterexamples} to the BNN verification problem.
Our work, however, goes well beyond current approaches to generating adverserial examples in that it does not only support debugging and retraining purposes. 
Instead, our verification algorithm establishes formal correctness results for neural network-like structures.

\section{Conclusions}~\label{sec.conclusion}

\vspace{-4mm}

We are solving the problem of verifying BNNs by reduction to the problem of verifying combinatorial circuits, which itself is reduced to 
solving SAT problems.
Altogether, our experiments indicate that this hardware verification-centric approach, in connection with our BNN-specific transformations and optimizations, scales to BNNs with thousands of inputs and nodes. 
This kind of scalability makes our verification approach attractive for automatically establishing correctness results 
at least for moderately-sized BNNs as used on current embedded devices.

Our developments for efficiently encoding BNN verification problems, however, might also prove to be useful 
in optimizing forward evaluation of BNNs.
In addition our verification framework may also be used for debugging and retraining purposes of BNNs; for example, for automatically generating 
adverserial inputs from failed verification attempts.

In the future we also plan to directly synthesize propositional clauses without the support of 3rd party tools such as \texttt{Yosys} in order to avoid extraneous transformations and repetitive work in the synthesis workflow.
Similar optimizations of the current verification tool chain should result in substantial performance improvements.
It might also be interesting to investigate incremental verification techniques for  BNN, since weights and structure of these learning networks might adapt and change continuously.

Finally, our proposed verification workflow might be extended to synthesis problems, such as synthesizing bias terms in BNNs without sacrificing 
 performance or for synthesizing weight assignments in a property-driven manner. These kinds of synthesis problems for BNNs are reduced to 2QBF problems, which are satisfiability problems with a top level exists-forall quantification. The main 
challenge for solving these kinds of synthesis problems for the typical networks encountered in practice is, again, scalability.

\vspace{3mm}
\noindent{\textbf{Acknowledgments}} 
\vspace{-2mm}
\begin{description}
	\item[(Version 1)] We thank Dr. Ljubo Mercep from Mentor Graphics for indicating to us some recent results on quantized neural networks, and Dr. Alan Mishchenko from UC Berkeley for his kind support regarding~$\texttt{ABC}$. 
	
	\item[(Version 2)] We additionally thank Dr. Leonid Ryzhyk from VMWare to indicate us their work on  efficient SAT encoding of individual neurons\@. We further thank Dr. Alan Mishchenko from UC Berkeley for sharing his knowledge  regarding sorting networks. 
\end{description}

\bibliographystyle{abbrv}

\section*{Appendix - Proof of Theorems}

\vspace{3mm}

\setcounter{exe}{0}

\begin{exe}
	The problem of BNN safety verification is NP-complete. 
\end{exe}

\proof Recall that for a given BNN and a relation $\phi_{risk}$ specifying the undesired property between the bipolar input and output domains of the given BNN, the {\em BNN safety verification problem} asks if there exists an input $\vec{a}$ to the BNN such that the risk property $\phi_{risk}(\vec{a},\vec{b})$ holds, where $\vec{b}$ is the output of the BNN for input~$\vec{a}$\@.

\vspace{1mm}
\noindent (NP) Given an input, compute the output and check if $\phi_{risk}(\vec{a},\vec{b})$ holds can easily be done in time linear to the size of BNN and size of the property formula. 

\vspace{1mm}
\noindent (NP-hardness) The NP-hardness proof is via a reduction from 3SAT to BNN safety verification. Consider variables $x_1, \ldots , x_m$, clauses $c_1, \ldots, c_n$ where for each clause $c_j$, it has three literals $l_{j_1}, l_{j_2}, l_{j_3}$. We build a single layer BNN with inputs to be $x_0=+1 $ (constant for bias), $x_1, \ldots, x_m, x_{m+1}$ (from CNF variables), connected to $n$ neurons. 

For neuron $n^{1}_j$, its weights and connection to previous layers is decided by clause~$c_j$. 

\begin{itemize}
	\item If $l_{j_1}$ is a positive literal $x_i$, then in BNN create a link from $x_i$ to neuron~$n^{1}_j$ with weight $-1$. If $l_{j_1}$ is a negative literal $x_i$, then in BNN create a link from $x_i$ to neuron~$n^{1}_j$ with weight $+1$. Proceed analogously for $l_{j_2}$ and $l_{j_3}$.  
	\item Add an edge from $x_{m+1}$ to $n^{1}_j$ with weight $-1$.
	\item Add a bias term $-1$.
\end{itemize}

For example, consider the CNF having variables $x_1, \ldots, x_6$, then the translation of the clause $(x_3 \vee \neg x_5 \vee x_6)$ will create in BNN the weighted sum computation $(-x_3+x_5-x_6)-x_7-1$.

Assume that $x_7$ is constant $+1$, then if there exists any assignment to make the clause $(x_3 \vee \neg x_5 \vee x_6)$ true, then by interpreting the $\sig{true}$ assignment in CNF to be~$+1$ in the BNN input and $\sig{false}$ assignment in CNF to be~$-1$ in the BNN input, the weighted sum is at most~$-1$, i.e., the output of the neuron is $-1$. Only when $x_3 = \sig{false}$, $x_5 = \sig{true}$ and $x_6 = \sig{false}$ (i.e., the assignment makes the clause unsatisfiable), then  the weighed sum is~$+1$, thereby setting output of the neuron to be~$+1$.

Following the above exemplary observation, it is easy to derive that 3SAT formula is satisfiable  \emph{iff} in the generated BNN, there exists an input such that the risk property $\phi_{risk}:= (x_{m+1}= +1 \rightarrow (\bigwedge^{n}_{i=1} x^{(1)}_{i} = -1))$ holds. It is done by interpreting the 3SAT variable assignment $x_i := \sig{true}$ in CNF to be assignment $+1$ for input  $x_i$ in the BNN, while interpreting  $x_i := \sig{false}$ in 3SAT to be $-1$ for input  $x_i$ in the BNN. \qed

\begin{exe}[Hardness of factoring optimization]
	The $k$-factoring optimization problem, even when $k=1$,  is NP-hard.
\end{exe}

\begin{figure}
	\centering
	\includegraphics[width=0.65\columnwidth]{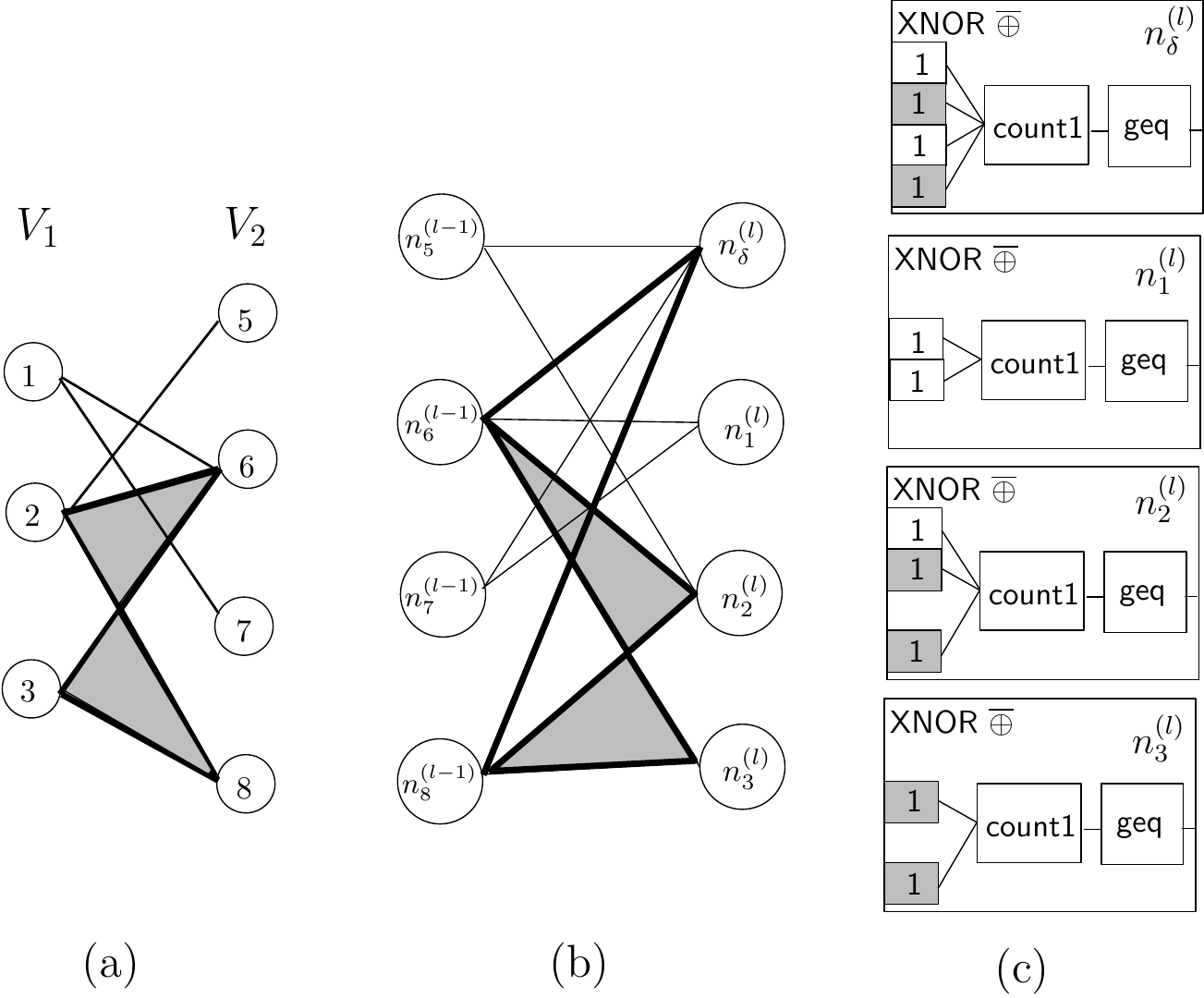}	
	\caption{From bipartite graph (a) to BNN  where all weights are with value~$1$ (b), to optimal factoring (c).}
	\label{fig:proof}		
\end{figure}

\proof 
The proof proceeds by a polynomial reduction from the problem of finding maximum edge biclique in bipartite graphs~\cite{peeters2003maximum}\footnote{ 
	Let $G = (V_1, V_2, E)$ be a bipartite graph with vertex set $V_1 \uplus V_2$ and edge set $E$ connecting vertices in $V_1$ to vertices in $V_2$. A pair of two disjoint subsets $A\in V_1$ and $B\in V_2$ is called a \emph{biclique} if $(a, b) \in E$ for all $a\in A$ and $b\in B$. Thus, the edges $\{(a, b)\}$ form a complete bipartite subgraph of $G$. A biclique $\{A; B\}$ clearly has $|A|\cdot|B|$ edges.}. 
Given a bipartite graph $G$, this reduction is defined as follows.
\begin{enumerate}
	\item For $v_{1\alpha}$, the $\alpha$-th element of $V_1$, create a neuron $n^{(l)}_{\alpha}$. 
	\item Create an additional neuron $n^{(l)}_{\delta}$
	\item  For $v_{2\beta}$, the $\beta$-th element of $V_2$, create a neuron $n^{(l-1)}_{\beta}$.
	\begin{itemize}
		\item Create weight $w^{(l)}_{\beta\delta} = \sig{1}$.
		\item If $(v_1, v_2) \in E$, then create $w^{(l)}_{\beta\alpha} = \sig{1}$.
	\end{itemize}
\end{enumerate}
This construction can clearly be performed in polynomial time. 
Figure~\ref{fig:proof} illustrates the construction process. It is not difficult to observe that $G$ has a
maximum edge size~$\kappa$ biclique $\{A; B\}$ iff the neural network at layer $l$ has a factoring $(I, J)$ whose saving equals $(|I|-1)\cdot |J| = \kappa$. 
The gray area in Figure~\ref{fig:proof}-a shows the  structure of maximum edge biclique $\{\{1,2\};\{6,8\}\}$. For Figure~\ref{fig:proof}-c, the saving is $(|\{n^{(l)}_{\delta}, n^{(l)}_{2}, n^{(l)}_{3}\}|-1)\cdot 2 = 4$, which is the same as the 
edge size of the biclique.
\qed


\vspace{5mm}

The following inapproximability result shows that even having an approximation
algorithm for the $k$-factoring optimization problem is hard.

\begin{exe}
	Let $\epsilon > 0$ be an arbitrarily small constant. If there is a PTAS
	for the $k$-factoring optimization problem, even when $k=1$, then there
	is a (probabilistic) algorithm that decides whether a given SAT
	instance of size $n$ is satisfiable in time~$2^{n^{\epsilon}}$.	
\end{exe}

\proof

We will prove the Theorem by showing that a PTAS for the $k$-factoring
optimization problem can be used to manufacture a PTAS for MEB. Then the result
follows from the inapproximability of MEB assuming the exponential time
hypothesis~\cite{ambuhl2011inapproximability}.

Assume that $\mathcal{A}$ is a $\rho$-approximation algorithm for the
$k$-factoring optimization problem. We formulate the following algorithm
$\mathcal{B}$:\\

\noindent \textbf{Input:} MEB instance $M$ (a bipartite graph $G=(V, E)$)

\noindent \textbf{Output:} a biclique in $G$
\begin{enumerate}
	\item perform reduction of proof of Theorem~1 to obtain $k$-factoring
	instance $F:=\rm{reduce}(M)$
	\item factoring $(I, J):=\mathcal{A}(F)$
	\item return $(I\setminus\{n^{(l)}_{\delta}\}, J)$
\end{enumerate}
{\tiny Remark: step 3 is a small abuse of notation. It should return the
	original vertices corresponding to these neurons.}

\noindent Now we prove that $\mathcal{B}$ is a $\rho$-approximation algorithm for MEB:
Note that by our reduction two corresponding MEB and $k$-factoring instances
$M$ and $F$ have the same optimal value, i.e.,
$\textsc{Opt}(M)=\textsc{Opt}(F)$.

In step 3 the algorithm returns $(I\setminus\{n^{(l)}_{\delta}\}, J)$. This is
valid since we can assume w.l.o.g.\ that $I$ returned by $\mathcal{A}$ contains
$n_{\delta}^{(l)}$. This neuron is connected to all neurons from the previous
layer by construction, so it can be added to any factoring. The following
relation holds for the number of edges in the biclique returned by
$\mathcal{B}$:
\begin{subequations}
	\begin{eqnarray}
	\|I\setminus\{n^{(l)}_{\delta}\}\|\cdot \|J\| & = & (\|I\|-1)\cdot
	\|J\| \\
	& \geq & \rho \cdot \textsc{Opt}(F) \label{eq:assumption}\\
	& = & \rho \cdot \textsc{Opt}(M) \label{eq:reduction_equality}
	\end{eqnarray}
	\label{eq:proof:inapprox}
\end{subequations}

The inequality in step~\eqref{eq:assumption} holds by the assumption that
$\mathcal{A}$ is a $\rho$-approximation algorithm for $k$-factoring and
\eqref{eq:reduction_equality} follows from the construction of our reduction.
Equations~\eqref{eq:proof:inapprox} and the result
of~\cite{ambuhl2011inapproximability} imply Theorem~2.

\qed

\end{document}